%% file: deform_source1.tex
\documentclass[aps,reprint]{revtex4-2}
\usepackage{amsmath}

\usepackage{adjustbox}
\usepackage{tikz}
\usepackage{float}
\usepackage{tabularx}
\usepackage{soul}
\usepackage[normalem]{ulem}
\usepackage{url}
\usepackage{graphicx}
\usepackage{xr-hyper}
\usepackage{hyperref}
\usepackage{isotope}
\usepackage{mhchem}
\usepackage{xcolor}
\usepackage{tikz}
\usepackage{float}
\usepackage{tabularx}
\usepackage{makecell}
\usepackage{amsmath}
\usepackage{amssymb}
\usepackage{sidecap}
\usepackage[T1]{fontenc}
\usepackage[utf8]{inputenc}
\usepackage{xcolor}
\definecolor{pastelgray}{rgb}{0.81, 0.81, 0.77}
\definecolor{beaublue}{rgb}{0.9, 0.9, 0.93}
\usepackage{booktabs}

\begin{document}
\title{Proton Femtoscopy in Heavy-Ion Collisions: From Transport Theory to Source Imaging}

\author{Pierre Nzabahimana$^{1,2}$, Pawel Danielewicz$^{2}$, and Giuseppe Verde$^{3}$}
\affiliation{$^{1}$Los Alamos National Laboratory, Los Alamos, NM 87545}
 \affiliation{$^{2}$Facility of Rare Isotope Beams and Department of Physics and Astronomy, \\
 	Michigan State University, East Lansing, Michigan 48824, USA}
 \affiliation{$^{3}$Istituto Nazionale di Fisica Nucleare,Catania, Sicily, Italy}

\begin{abstract}
\input{abstract}

\end{abstract}
\maketitle
\section{introduction}
Transport models are used to extract information about the nuclear Equation of State (EoS) (see~\cite{WOLTER2022103962} and Refs.\ therein) from heavy-ion collision data.  Results from extensive studies performed with the Boltzmann-Uehling-Uhlenbeck model (BUU), the Quantum Molecular Dynamics model (QMD), and the Antisymmetrized Molecular Dynamics Model (AMD) can be found in the literature \cite{DANIELEWICZ2000375, WOLTER2022103962, AMD}.  These models solve nuclear transport equations governed by interaction terms directly linked to the EoS.  The results of their simulations are compared with experimentally measured observables to infer the EoS and other transport properties used as input parameters.  More recently, the studies of the N/Z-asymmetry dependence of the EoS have become increasingly important because of their potential implications on the properties of the interior of neutron stars (see \cite{Sorensen2024} and Refs.\ therein).  The accelerator facilities providing beams with a range of neutron-proton asymmetries and the transport simulations have therefore provided tools to study nuclear matter of astrophysical relevance under laboratory-controlled conditions. 

Understanding the complex dynamics of Heavy-Ion Collisions (HIC) requires theoretical and experimental convergence around observables that probe the space-time features of the collision process.  In this context, particle-particle correlations have been widely used~\cite{VerdeBuu, verde_imaging_2002}, and transport models have provided a framework for understanding the properties of nuclear matter that they may help to study.  In this paper, we use the pBUU transport approach, developed by Danielewicz and Bertsch~\cite{danielewicz_production_1991}, describing the time evolution of a single-particle phase-space distribution under the influence of mean fields and nucleon-nucleon collisions~\cite{BERTSCH1988189, VerdeBuu, WOLTER2022103962}.  


The quantity linking BUU models \cite{VerdeBuu,bauer1993particle} with correlation data is the so-called two-particle source function~$S(r)$, that is, the probability density of emitting a pair of particles at a relative distance $r$.  Since the two particles in the pair are not emitted simultaneously, the relative distance $r$ is evaluated at the time when the second particle is emitted.  Following the so-called Koonin-Pratt equation \cite{koonin1977proton, pratt1984pion, brown1997imaging}, one can fold the square of a final-state wave function calculated theoretically with the source from transport to arrive at a correlation function to be compared with the data. 

Since BUU involves several adjustable parameters, one can likely learn from correlation-function data by comparing the source function obtained from BUU with that from imaging methods \cite{verde_imaging_2002}.  Recently, a new imaging approach called "deblurring" has been introduced in Refs.~\cite{Pierre,nzabahimana2023source,danielewicz2022deblurring}.
 The parameters that yield a source function that better conforms to the outcomes of imaging and deblurring approaches can be used to constrain the nuclear equation of state (EoS) and other important transport properties.
For example, in Ref.~\cite{VerdeBuu}, the source function calculated with BUU simulations was compared to the sources imaged from experimental correlation functions.  The two-proton correlation was found to be sensitive to the details of the in-medium nucleon-nucleon cross-section, $\sigma_{NN}$. 

To learn about the two-particle emitting source function from correlations, one needs to solve the Koonin-Pratt (KP) equation \cite{koonin1977proton, pratt1984pion, brown1997imaging},
\begin{eqnarray}
C(\mathbf{q}) = \int \, d^3 r  \, K(\mathbf{q},\mathbf{r}) \, S(\mathbf{r}) \, .
\label{KPeqn}
\end{eqnarray}
In this equation, the correlation function, $C(\mathbf{q})$, is expressed as a function of the relative momentum for the two equal-mass particles, $\mathbf{q}=(\mathbf{p}_{1}-\mathbf{p}_{2})/2$, with the momentum vectors of $\mathbf{p}_1$ and $\mathbf{p}_2$ of the two particles in the center of mass.  The right-hand side of the equation contains the integral over relative distance of the product between the kernel function $K(\mathbf{q},\mathbf{r})$ and the source function~$S(\mathbf{r})$, namely the probability density of the pair emitted at relative distance $\mathbf{r}$. In the equation, the relative distance $\mathbf{r}$ is calculated when the second particle is emitted~\cite{koonin1977proton, pratt1984pion, BROWN1997252}.  The kernel is the square modulus of the pair relative wave function with the outgoing boundary condition of the relative momentum $\mathbf{q}$ imposed at the detectors.  For details on how the KP formula can be derived, we refer the reader to Ref.~\cite{danielewicz_formulation_1992}.  
On the other hand, experimentally, two-particle correlations can be measured in heavy-ion collisions as the ratio between two-proton coincidence yields, $Y_{coinc}(\mathbf{p}_{1},\mathbf{p}_{2})$ and the product of single particle proton spectra, $Y(\mathbf{p}_{1})\cdot Y(\mathbf{p}_{2})$, where $\mathbf{p}_1$ and $\mathbf{p}_2$ are the momenta of the protons~\cite{verde2006correlations,boal1990}:
\begin{eqnarray}
C({\bf q})\equiv 1+R({\bf q}) = c\, \frac{Y_{coinc}({\bf p}_{1},{\bf p}_{2})}{Y({\bf p}_{1})\, Y({\bf p}_{2})} \, .
\label{eq:C_exp}
\end{eqnarray}
Here, $c$ is a constant to be chosen such that $R(q)=0$ at large $q$-values.   Due to statistics and limitations in angular coverage, most often in the literature, the possible dependence of the correlation on the $\bf q$ direction was not pursued, with the data averaged over the directions of $\bf q$ relative to the beam, reaction plane, or $\bf P$.  In data interpretations, source functions have typically been parameterized as Gaussians due to their simple form~\cite{verde2006correlations}.  Then, the best fits to the correlation data, following the KP equation \eqref{eq:C_exp}, yielded the widths of Gaussian sources~\cite{goldhaber_influence_1960, PhysRevC.33.549, pratt1984pion, boal90, verde_imaging_2002}. 


 In this paper, we study two-proton angle-averaged source functions extracted from pBUU simulations of $^{36}$Ar+$^{45}$Sc, $^{40}$Ca+$^{40}$Ca, and $^{48}$Ca+$^{48}$Ca collisions at E/A=80 MeV, and assess the effect of protons emitted from long-lived emission processes on the correlation function. 
 This is achieved by comparing the pBUU source functions with source profiles extracted from experimentally measured two-proton correlation functions, analyzed with the {\it deblurring method}.  This analysis technique was shown to recover input sources~\cite{nzabahimana2023source} in tests of its numerical robustness, and it represents a step forward in studying correlations without a priori assumptions about the functional form of the emitting source function, which is most commonly Gaussian~\cite{nzabahimana2023source,tam2025, Xu2025}. The experimental data were collected in separate experiments~\cite{Handzy, Henzl} at the National Superconducting Cyclotron Laboratory (NSCL) at Michigan State University, which was later incorporated into the Facility for Rare Isotope Beams.

The performed pBUU simulations do not account for long-lived emission processes, such as secondary decays of primary fragments, that mainly contribute to the source function at large $r$-values. In this region of the source function, the mutual Coulomb repulsion dominates the final-state interactions of proton pairs accounted for in the relative wavefunction. The most pronounced impact on the correlation function, from protons where one or both particles are emitted via these long-lived processes, is observed at small values of $q$, such as $q \lesssim 10 \, \text{MeV}/c$, modifying the anticorrelation there. When data extend into that $q$-region, they can inform about the source through deblurring, and we can isolate the contributions of long-lived processes more clearly by comparing pBUU sources with those inferred from data through deblurring.  We introduce a parameterization of long-lived process contributions for use when no deblurring is employed.  We demonstrate the separation of source-tail for the reactions at $E/A = 80$~MeV$:\ ^{36}\mathrm{Ar} + ^{45}\mathrm{Sc}$, already briefly discussed in~\cite{nzabahimana2025source}, and $^{40}\mathrm{Ca} + ^{40}\mathrm{Ca}$ and $^{48}\mathrm{Ca} + ^{48}\mathrm{Ca}$.  In the latter two cases, we examine the tail-core separation of the source for different angular-momentum cuts in proton-pair emission.


The manuscript is organized as follows. Section \ref{sect2} presents a theoretical calculation of two-proton correlations based on pBUU transport-model simulations.  Section~\ref{chap63} is dedicated to the extraction of sources from correlation data, with the $^{36}$Ar + $^{45}$Sc reaction as an example, and comparison of the inferred source to transport. Section~\ref{chapV} focuses on reconciling inferences from transport and data for the $^{36}$Ar + $^{45}$Sc case by adding a parametrized long-term emission source to the transport source.  Next, Sec.~\ref{sect5} turns to the sources in Ca + Ca reactions and changes in the sources due to the reactions' angular dependence. We summarize the results of our work in Section~ \ref{sum}.

\section{Two-proton correlation functions from BUU Model \label{sect2}}
The BUU model is a semi-classical one-body transport model extensively used to describe the dynamics of heavy-ion collisions at intermediate energies. In this model, all one-body observables for a system are determined by single-particle phase-space densities $f=f(\mathbf{p,r},t)$ that evolve self-consistently with time, according to the Boltzmann equation,
\begin{eqnarray}
\Big(\frac{\partial}{\partial t}+\mathbf{v}_\mathbf{p} \cdot \nabla_\mathbf{r} -\nabla_\mathbf{r} U \cdot \nabla_\mathbf{p}\Big)f=I_{col} \, ,
\label{BUU}
\end{eqnarray}
 with finer technical details suppressed, such as relativity, little relevant for the studied beam energies.
The left-hand side of Eq.~\eqref{BUU} accounts for the motion of the particle at velocity $\mathbf{v}_\mathbf{p}=\nabla_\mathbf{p} \epsilon $ subject to force $\mathbf{F}-\nabla _\mathbf{r} U$.  Here, $\mathbf{p}$ and $\epsilon$ are the momentum and single-particle energy, $\epsilon = \epsilon_{kin}+U$, respectively. On the right-hand side, $I_{col}$ is the collision term that accounts for nucleon-nucleon collisions:
\begin{eqnarray}
    I_{col}(\mathbf{p}_1,\mathbf{r},t)&=&\frac{g}{h^3}\int d^3\mathbf{p}_2\int d\Omega \, \frac{d\sigma_{12}}{d \Omega} \, v_{12}  \nonumber\\
    & &\times \Big[\hat{f}_1\hat{f}_2f_{1'}f_{2'}-f_1f_2\hat{f}_{1'}\hat{f}_{2'}\Big]  \nonumber \\
    &=& \Sigma(\mathbf{p}_1,\mathbf{r},t)  - \Gamma \cdot f(\mathbf{p}_1,\mathbf{r},t) \, .
    \label{eq:Icol}
\end{eqnarray}
Here, $\hat{f}$  is the statistical factor equal to $1- f$ for fermions, $g=2$ is spin degeneracy, and $\frac{d\sigma_{12}}{d \Omega}$ is the differential cross section for scattering to angle $\Omega$. Indices 1 and 2 refer to the initial momenta in a collision, $\mathbf{p}_1$ and $\mathbf{p}_2$, and the primed indices refer to the final momenta.  The relative velocity $v_{12}$ is related to the nucleon-nucleon (NN) relative momentum $\mathbf{q}_{12}$, with $\mu \mathbf{v}_{12} \equiv \mu  (\mathbf{v}_1-\mathbf{v}_2) = \mathbf{q}_{12}$, and for equal-mass particles, $\mathbf{q}_{12} = (\mathbf{p}_1 - \mathbf{p}_2)/2$ in any reference frame, with the same convention as in the context of the low relative velocity correlation functions.  The NN collisions take place at conserved total momentum for the pair $ \mathbf{P}_{12}=\mathbf{p}_1+\mathbf{p}_2$. In the r.h.s.\ of Eq.~\eqref{eq:Icol}, we introduced the damping rate $\Gamma$, as a factor multiplying $f_1$ in $I_{col}$, and the feeding rate $\Sigma$, as a remnant.  The formal solution for any $f$, when starting at a time $t_0$, may be written as 
\begin{equation}
    \begin{split}
    f(\mathbf{p}(t),\mathbf{r}(t),t) = & \int_{t_0}^t dt_\sigma \, \text{e}^{ -\int_{t_\sigma}^t dt' \, \Gamma(\mathbf{p}(t'),\mathbf{r}(t'),t') } \\ & \times \Sigma (\mathbf{p}(t_\sigma),\mathbf{r}(t_\sigma),t_\sigma)
    \\ & + \text{e}^{ -\int_{t_0}^t dt' \, \Gamma(\mathbf{p}(t'),\mathbf{r}(t'),t') } \, f(\mathbf{p}_0,\mathbf{r}_0,t_0) \, .
    \end{split}
    \label{eq:Solution}
\end{equation}
Here, $\mathbf{p}(t)$ and $\mathbf{r}$ satisfy phase-space trajectory equations, $d\mathbf{p}/dt = \mathbf{F}$ and $d\mathbf{r}/dt=\mathbf{v}$.  Due to collisions, particles moving around a phase space trajectory $(\mathbf{p}(t),\mathbf{r}(t))$ are depleted, but the scattering also populates the surroundings of that trajectory, at the rate $\Sigma$.  The physics inherent in Eq.~\eqref{eq:Solution} may be expected to apply even when the rates provided by the Boltzmann equations are deficient.

To solve the set of Boltzmann equations in practice, we use the pBUU code, for which various technical details may be found in Ref.~\cite{DANIELEWICZ2000375}.  
The practical realization of transport is complex and requires several inputs.  These include an energy functional that yields Equation of State (EoS) in an equilibrium situation and single-particle energies $\epsilon$ for any conditions, with potentials $U$ that may exhibit momentum dependence.  The energy functional is employed both in preparing systems for collisions by solving the Thomas-Fermi equations~\cite{DANIELEWICZ2000375} and in transport.  In the calculations for this paper, we have employed a momentum dependence characterized by nucleon isoscalar effective mass $m^*=0.7 m$ at the Fermi momentum at normal density, in combination with a softer or stiffer version of incompressibility, $K=214 \, \text{MeV}$ or $280 \, \text{MeV}$.  We have also explored soft and stiff versions of the symmetry energy, with the potential term in the energy functional of the form ${\mathcal S}_{pot}(\rho)={\mathcal S}_0(\rho/\rho_0)^\gamma$, where $\rho_0=0.16~\text{fm}^{-3}$ is the nuclear-matter saturation density and $\gamma = 0.5$ and 1.6 for the soft and stiff case, respectively.  However, we found no significant impact of EoS on the correlations reported here.

In our pBUU simulations, the p-p source function can be directly calculated from the single-particle phase-space density $f=f(\mathbf{p,r},t)$, at a time $t$ late enough so that emission populating the particular momentum region fades out~\cite{VerdeBuu}:
\begin{eqnarray}
S^{BUU}(\mathbf{r}) = \frac{\int d^3R \, f(\frac{\mathbf{P}}{2},\mathbf{R}+\frac{\mathbf{r}}{2},t) \, f(\frac{\mathbf{P}}{2},\mathbf{R}-\frac{\mathbf{r}}{2},t)}{\big(\int d^3r' \, f(\frac{\mathbf{P}}{2},\mathbf{r}',t) \big)^2} \, ,
\label{SBUU}
\end{eqnarray}
In this equation, $\mathbf{r} = \mathbf{r}_1 - \mathbf{r}_2$ and $\mathbf{R}=(\mathbf{r}_1+\mathbf{r}_2)/2$ are, respectively, the relative and center of mass positions for a proton pair where each proton is at the same momentum equal to half of the total momentum, $\mathbf{P} = \mathbf{p_1}+\mathbf{p_2}$.
Having constructed the pBUU source function, the corresponding correlation function can be calculated following Eq.~\eqref{KPeqn_iso}:  
\begin{widetext}
\begin{eqnarray}
   C({\bf q})\equiv 1 +  R({\bf q}) =\int d^3 r \,  |\Psi_{pp}(\mathbf{q,r})|^2 \, \frac{\int d^3R \, f(\frac{\mathbf{P}}{2},\mathbf{R}+\frac{\mathbf{r}}{2},t) \, f(\frac{\mathbf{P}}{2},\mathbf{R}-\frac{\mathbf{r}}{2},t)}{\big(\int d^3r' \, f(\frac{\mathbf{P}}{2},\mathbf{r}',t)\big)^2} \, .
   \label{CF}
\end{eqnarray}
\end{widetext}

Eqs. ~\ref{SBUU} and~\eqref{eq:Solution} reveal the interplay between spatial and temporal features of emission encompassed in the rate $\Sigma$.  A spatial spread in $\Sigma$ around the pair velocity $\mathbf{v} = \mathbf{P}/2m$ contributes to a spread in~$S$ for the pair.  A temporal spread in $\Sigma$, also contributes to a spread in $S$, with details depending on the features of $\Sigma$ in \eqref{eq:Solution}.  Secondary decays of products that are fast compared to the motion at $v$ will broaden the tail of $S$ in all directions.  A prolonged emission from a compact source $\Sigma$ will give rise to a tail in $S$ along $\mathbf{v}_\Sigma - \mathbf{v}$, where $\mathbf{v}_\Sigma$ is the velocity of the region emitting the particles of which the correlations are studied.  An anisotropy in $C$, with a main axis along that direction, is then expected, too~\cite{PhysRevLett.75.4190}. 

Directional analyses can be useful for resolving potential space-time ambiguities in interpreting the shape of the source function~\cite{lisa2005femtoscopy}. However, important physics information is commonly extracted even in so-called angle-averaged studies, where the source function, the kernel, and the resulting correlation functions are integrated over the relative angle between total momentum and relative momentum, resulting in an angle-averaged version of the Koonin-Pratt equation: 
\begin{eqnarray}
    C(q) = 4 \pi \int dr \, r^2 K(q,r) \, S(r) \, .
    \label{KPeqn_iso}
\end{eqnarray}
In this work, we will focus on angle-averaged studies that are very useful for understanding femtoscopy in the context of models and its links to nuclear transport properties~\cite{VerdeBuu}.  Details on calculation of the angle-averaged kernel, $K(q,r)$, are provided in Refs.~\cite{brown1997imaging,BROWN1997252,brown1998optimized}. 

\begin{figure*}[!htb]
    \centering
    \includegraphics[scale=0.4]{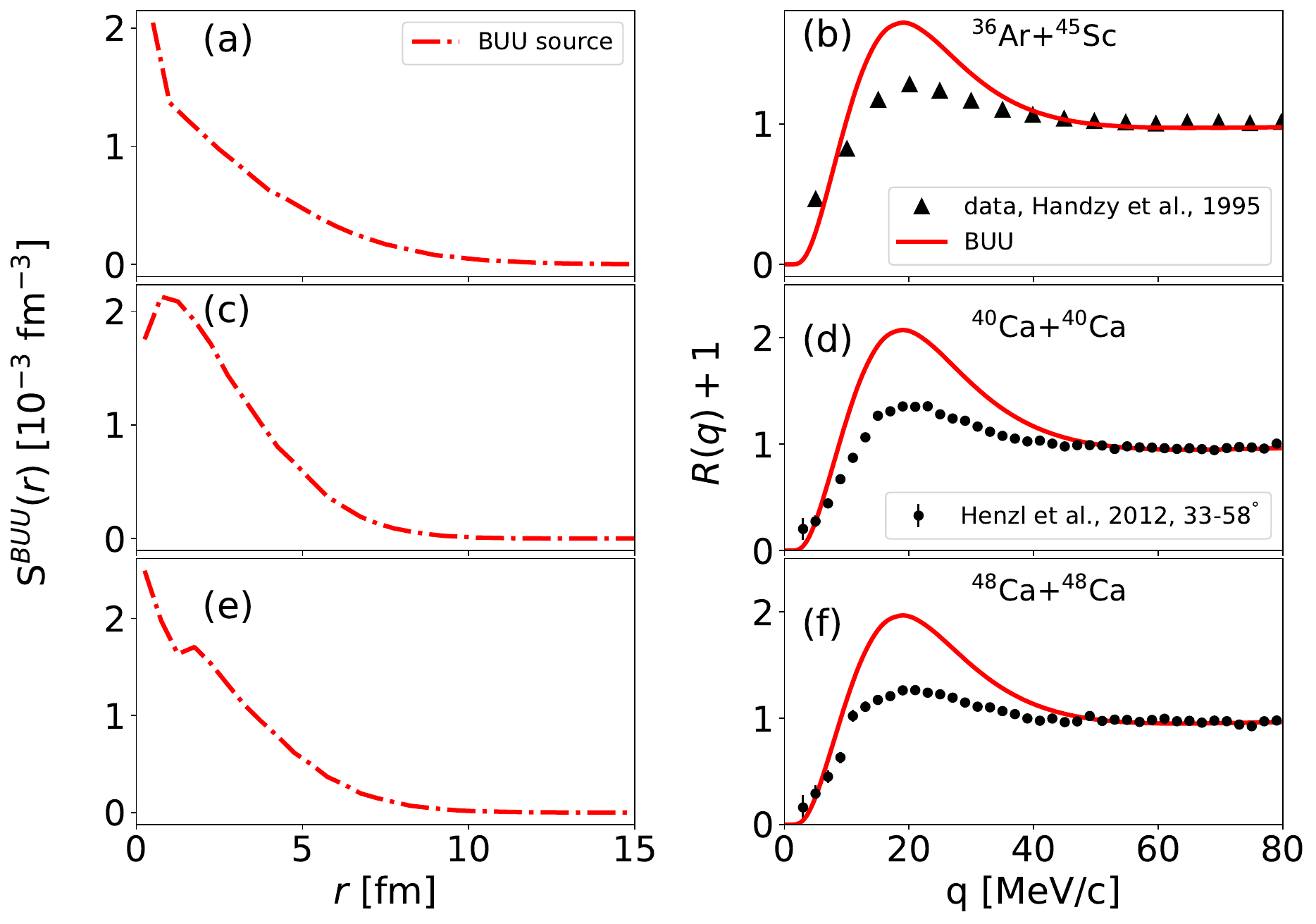}
    \caption{$pp$ relative source functions from BUU calculations are displayed on the left and the corresponding correlation functions on the right as lines, together with data as symbols, for different collisions. The top panels show results for the $^{36}\mathrm{Ar} + ^{45}\mathrm{Sc}$ collisions at $E/A=$ 80 MeV in the $pp$ total momentum bracket of $P=200$-$400 \, \text{MeV}/c$ and data from the work of Handzy {\em et al.}~\cite{Handzy}. The middle and bottom panels show, respectively, results from the $^{40}$Ca + $^{40}$Ca, and $^{48}$Ca +$^{48}$Ca collisions, at $E/A=$ 80 MeV, with the $pp$ pairs from the angular window of $33 < \theta \leq 58^{\circ}$  and total momentum bracket $P=500$-$640\, \text{MeV}/c$.  The data in the middle and bottom panels stem from the work of Henzl {\em et al.}~\cite{Henzl}.}
    \label{fig:CF_fun}
\end{figure*}

In pBUU simulations, two main stages of particle emission can be distinguished during the collision's evolution. There is an early stage of fast nucleon emission (often referred to as the pre-equilibrium stage), evolving over timescales below 50-100 fm/c, depending on beam energy, followed by another stage where particles are emitted over longer timescales~\cite{verde_imaging_2002}. The pre-equilibrium stage, associated with the early contact between the two reaction partners during the collision, makes a greater contribution to particles emitted at large transverse velocities. In this stage, particle pairs are emitted at similar velocities and at shorter time and space distances from each other. The following slower emission stage is characterized by particle pairs with lower transverse velocities and longer relative distances and emission times. 
Both emission stages contribute to the shape of the source function, though to different degrees depending on the particles' momentum region.  At any momentum, the pre-equilibrium emission alone contributes more strongly to the short-$r$ region of $S$. Instead, long-lived emissions from the second stage generally populate the source at large $r$-values~\cite{VerdeBuu}.

Fig.~\ref{fig:CF_fun} shows the proton–proton pBUU sources (left panels, dot-dashed lines) and correlation functions (right panels and solid lines) determined for simulations of collisions $^{36}\mathrm{Ar} + ^{45}\mathrm{Sc}$, $^{40}\mathrm{Ca} + ^{40}\mathrm{Ca}$ and $^{48}\mathrm{Ca} + ^{48}\mathrm{Ca}$ in $E/A = 80$~MeV and an impact parameter of $b = 1.9$~fm.  In the case of $^{36}\mathrm{Ar} + ^{45}\mathrm{Sc}$ collisions represented in panels (a) and (b), the calculations were performed for proton pair total momenta between 200 and 400 MeV/c, to compare to the experimental data from Ref.~\cite{Handzy} represented by triangles in panel (b).  In the case of $^{48}\mathrm{Ca} + ^{48}\mathrm{Ca}$ and $^{40}\mathrm{Ca} + ^{40}\mathrm{Ca}$, the proton pairs are selected from the total momentum bracket of $P=500$-$640\, \text{MeV}/c$ and the angular range of $33 < \theta \leq 58^{\circ}$ in the laboratory frame to allow for a direct comparison to the results of measurements at NSCL using the HiRA detector~\cite{Henzl} represented by solid circles in the  panels (d) and (f). 

As we have already indicated in Ref.~\cite{nzabahimana2025source}, for the particular system mass region and beam energy, on the scale of the encountered discrepancies between the simulation and data in Fig.~\ref{fig:CF_fun}, we find a relatively weak sensitivity of the source and correlation results to the details of the energy functional used in the simulations.  For this reason, we limit the presentation of the theoretical results to those obtained following the soft, momentum-independent EoS.


Source regions below about 5-8~fm from pBUU simulations in Fig.~\ref{fig:CF_fun} are dominated by the pairs of protons emitted at the earlier stages of the collision. These regions of source functions are expected to be qualitatively well described by the BUU approaches during the collision stage, where elementary nucleon-nucleon collisions and the EoS play a key role in the dynamics. The regions at large $r$-values, beyond 8-10~fm, in contrast, can be less reliable due to the lack of reliable description of long-lived decays that cannot be adequately modeled with a Boltzmann equation.  Physically, at least one member of the proton pair at large $r$ is likely to stem from secondary decays of excited secondary fragments.  Such decays are missing in BUU models and may therefore result in a relative overprediction of the strength of the source function at small $r$-values compared to large $ r$.



The main features of the proton-proton correlation function, evident both in simulations and in experimental data, reflect an interplay between interaction within the pair (both short-range strong attraction and long-range Coulomb repulsion) and proton Fermi statistics, resulting in an antisymmetrized two-proton wave function. All these effects are taken into account in the kernel function. The attractive nuclear interaction gives rise to a peak around $q$ = 20 MeV/c in the correlation function in the right panels of Fig.~\ref{fig:CF_fun}. The details in the shape of the peak significantly depend on a small~$r$ region of the source function~\cite{verde_imaging_2002}. Coulomb repulsion depletes the wave function at moderate relative distances, and the depletion hole increases in size as $q$ decreases, generally following an increase in radius for the classical Coulomb barrier. 

In the correlation functions on the right of Fig.~\ref{fig:CF_fun}, the Coulomb repulsion yields an anticorrelation at $q< 10 \, \text{MeV}/c$. The details of this drop-off towards $q=0$ depend on how much the source function extends past the classical Coulomb barrier for a given $q$.  Often, as illustrated by the triangles in panel~(b), experimental data points are not available at low $q$ due to the protons of a pair stacked in the same detection cell. This can be partly resolved with detectors that have higher granularity and angular resolution, such as the HiRA array used for the data shown as solid circles in panels (d) and (f)~\cite{Henzl}. In any case, the presence of large error bars, compared to other regions of the correlation function, usually results in a weak sensitivity of the data to the details of the tail of the source function at large $r$-values. 
On the theoretical side, due to the probabilistic nature of the source $S$, which integrates over $\mathbf{r}$ to 1, any uncertainties at large $r$ impact the overall strength of $S$ at short $r$.  Conversely, an analysis of the correlation at large~$q$ can provide information on the net fraction of pairs in $S$ at large $r$, because of the normalization condition, as will be discussed later.


The main difference observed between the simulated correlation functions and the data in Fig.~\ref{fig:CF_fun} is an overestimation of the height of the peak at 20 MeV/c by model simulations. The origin of this discrepancy and its implications for the source function will be investigated in the following sections by comparing the pBUU predictions with source functions extracted directly from the experimental correlation data using a novel imaging method.


%
\section{Imaging source through deblurring \label{chap63}} 
In this section, we apply the Richardson-Lucy (RL) algorithm as an inverse problem method for imaging the proton-proton source function from the measured p-p correlation function. As discussed in our earlier publications, \cite{nzabahimana2023source,tam2025,nzabahimana2025source}, to implement the RL algorithm for the source function, the Koonin-Pratt equation \eqref{KPeqn_iso} needs to be rewritten in a discretized form:
\begin{equation}
    C_i=\sum_j K_{ij} \, \mathcal{S}_j \,  .
    \label{RLr}
\end{equation}
where $K_{ij}$ and $S_j$ are the discretized kernel and source function, respectively, with details provided in Ref.~\cite{nzabahimana2023source}.

\begin{figure*}[!htb]
    \centering
    \includegraphics[scale=0.4]{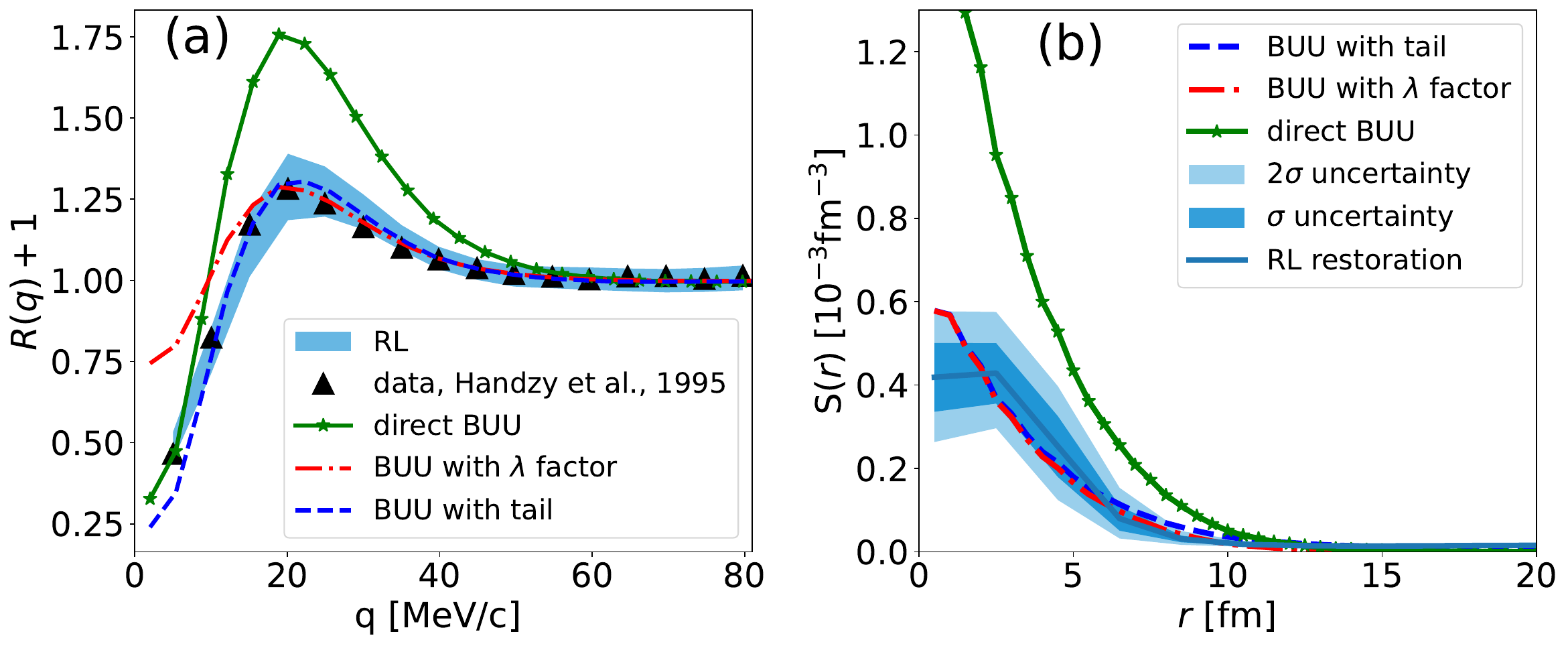}
    \caption{ Proton-proton correlation (a) and source (b) in central  $^{36}\mathrm{Ar} + ^{45}\mathrm{Sc}$ 
    collisions at $80 \, \text{MeV/nucleon}$, within the total momentum range of (200-$400) \, \text{MeV}/c$.  The triangles in (a) represent the correlation data of Ref.~\cite{Handzy}.  The shaded regions in~(b), with a central line, represent the results from deblurring the data in~(a).  The correlation results corresponding to the $2 \sigma$ uncertainty are indicated as a shaded region in~(a).  Finally, pBUU results are shown in both panels: the direct version and the version with a tail and corresponding renormalization of the direct contribution.} 
    \label{fig:SF_corr}
\end{figure*}

The RL algorithm determines the distribution $S$ when $C$ and $K$ are known.  To arrive at the RL strategy, a backward relation between $C$ and $S$ is invoked, which involves a conditional probability $\Tilde{K}$ that is complementary to $K$.  The fulfillment of a Bayesian relation involving $K$, $\Tilde{K}$, and $S$ is arrived at through iterations~\cite{richardson1972bayesian, nzabahimana2023particle},
\begin{equation}
    S^{(\mathfrak{n}+1)}_j = \mathcal{M}^{(\mathfrak{n})}_j \, S^{(\mathfrak{n})}_j \,  ,
    \label{eq:RL}
\end{equation}
where
\begin{equation}
    \mathcal{M}^{(\mathfrak{n})}_j =  \frac{\sum_i \alpha_i \frac{C_i}{C_i^{(\mathfrak{n})}} \, K_{ij} }{\sum_{i'} \alpha_{i'} K_{i'j}} \,  .
    \label{eq:RL1}
\end{equation}
Here, $\mathfrak{n}$ is the iteration index, $\mathcal{M}^{(\mathfrak{n})}_j$ is an amplification factor, and $C^{(\mathfrak{n})}$ is the prediction of the correlation function in the $\mathfrak{n}^\mathfrak{th}$ iteration:
The weights $\alpha$ on the r.h.s.\ of Eq.~\eqref{eq:RL1} can serve as importance factors and may be chosen as inverse errors or weights that amplify the importance of the region of low relative momentum where $C$ can really tell about~$S$.

Here, we next use the RL algorithm to extract the proton-proton source function from the experimental correlations reported in Refs.~\cite{Handzy, Henzl}. 
In Fig.~\ref{fig:SF_corr}(a), we show the proton-proton correlation functions for the reaction $^{36}\mathrm{Ar} + ^{45}\mathrm{Sc}$ at 80 MeV/nucleon and a total pair momentum range of 200–400 MeV/c. The triangles correspond to the experimental data. The deblurred source functions, obtained with the RL algorithm described in Eq.~\eqref{eq:RL}, are displayed as shaded blue bands in panel~(b). These deblurred source functions are then used in the KP equation, Eq.~\eqref{KPeqn_iso}, to obtain the deblurred correlation function shown as the corresponding blue-shaded band in panel~(a). 

We employ an error resampling technique to quantify the uncertainties in these reconstructed source functions and deblurred correlations (see Ref.~\cite{Pierre} for details). The results from this error resampling on the measured correlations are shown as bands: the dark and light blue bands in panel (b) correspond to the intervals $\sigma$ and 2$\sigma$, respectively. The solid blue line denotes the mean source function of 1000 samples. Ideally, measured experimental errors would be used for resampling; however, the data extracted from~\cite{Handzy} lack detailed error information. Consequently, the uncertainties are approximated using a Gaussian distribution, $\mathcal{N}(0, \sigma)$, with $\sigma = 0.045$ for relative momenta $q \geq 15$ MeV/c and $\sigma = 0.1$ for $q \leq 15$ MeV/c, where larger uncertainties are expected. Additionally, the blue band in the correlation functions in panel (a) represents the uncertainty of 2$\sigma$ associated with the source function.

As mentioned above, measured correlations are affected by detector resolutions, particularly at low values of $q$. To account for these resolution effects in our calculation, the proton-proton kernel is smeared in $q$ by folding the original kernel with a Gaussian in $q$ of width $\sigma_q=$2.8 MeV/c~\cite{nzabahimana2023source} and references within.

In the context of statistical uncertainties introducing short-wavelength variations in the input data, it is essential to discuss possible instability in the RL and other inversion methods for integral relations, such as the KP equation.  A forward relation, such as the KP relation, tends to suppress the impact of short-wavelength components, and a reverse relation tends to amplify them, including those spuriously induced by finite measurement statistics, raising the potential for an instability.  Generally, the positive definiteness of the restored source $S$ stabilizes the inversion~\cite{mambaAnalysisIterativeDeblurring2025}.  In addition, we employ the Total Variation regularization, as described in our previous studies, see Refs.\cite{Pierre, danielewicz2022deblurring, nzabahimana2023source}, with a small regularization factor of 0.0005. 
Importantly, recent work~\cite{tam2025} has shown that maximum entropy regularization is also effective in extracting the proton-proton source function from experimental correlations using the RL algorithm.

The reconstructed source shown in Fig. 2(b) exhibits a core component that rapidly decreases with $r$, together with a non-negligible tail extending beyond approximately 6–8 fm. Although the deblurring procedure successfully reproduces the measured correlation function, the physical origin of this long-range component is not immediately evident from the imaging analysis alone. In the next section, we compare the reconstructed source with the pBUU transport calculations and assess whether the observed tail can be associated with emission processes not included in the transport description.
\\


\section{pBUU study of the two-proton emitting source \label{chapV}} 

We next compare the source function, obtained with the RL algorithm and described in the previous sections, with the predictions of the pBUU simulations. 

The solid green line connecting stars in Fig.~\ref{fig:SF_corr}(b) represents the source function of pBUU, $S^{BUU}(r)$, calculated using Eq.~\eqref{SBUU}.
The corresponding correlation function, obtained by inserting this pBUU source into the KP equation, Eq.~\eqref{CF}, is shown in panel (a), as a green solid line connecting stars, together with the data as triangles.  The height of the peak at 20 MeV/c that reflects the strength of the core part of the calculated source, well higher than the data, demonstrates the overestimation of the fast-emission contribution to the source already discussed in Sect.~\ref{sect2}. Comparison between reconstructed and transport-generated source functions reveals that pBUU reproduces the falloff of the source at moderate relative distances, up to $\approx 8 \, \text{fm}$, where prompt proton emission dominates. However, at higher distances the pBUU source falls off too quickly compared to the reconstructed source.

One way of correcting for the overestimation of the short-range source strength and the corresponding excessive strength in the correlation is source renormalization. We renormalize the pBUU source with a multiplier factor~$\lambda$, following the procedure described in~\cite{VerdeBuu} and in Section II. As discussed before, the need for the renormalization can be tied to the lack of protons from secondary decays in pBUU. As the source function integrates to unity, this lack of late emissions results in an increasing strength of the source at short relative distances $r$. The resulting renormalized pBUU source is shown as a dot-dashed red line in Fig.~\ref{fig:SF_corr}(b). After adjustment of the strength of the pBUU source with $\lambda = 0.38$, its shape generally agrees with the source reconstructed from the data at short relative distances $r < 6 \, \text{fm}$.  The corresponding renormalized correlation function is shown as a red dashed-dotted line in panel (a) of  Fig.~\ref{fig:SF_corr}. Although the transport-derived correlation function now agrees with the data above $q \approx 18 \, \text{MeV}/c$, it does not follow the data below. Notably, early $pp$ correlation data might miss lower $q$-data results due to detector granularity and statistics \cite{boal1990}.

Using source imaging as leverage, we introduce here a strategy that allows us to compare transport predictions to nucleon-nucleon correlation data, despite the transport theory missing secondary decays, while simultaneously constraining that missing contribution to nucleons.
Specifically, we complement the transport-model source with a component in the source function that represents pairs where at least one member stems from late decays that are generally absent from transport approaches:
\begin{equation}
    S(r)= \lambda \, S_{BUU}(r) + (1-\lambda)\, S_{long}(r) \, ,
    \label{eq:long}
\end{equation}
where $S_{BUU}$ is calculated in the transport model and $S_{long}$ is parametrized; the two contributions differ with the range, both are normalized to 1, and $\lambda$ ensures normalization for $S$.  Such source decomposition can open up the possibility of testing the physics in transport models that these models should faithfully describe, while also providing greater access to the missing physics when compared to data.  In this work, we employ a phenomenological form for $S_{long}$, dropping exponentially with $r$, with just one fall-off parameter $B$~\cite{nzabahimana2025source}:
\begin{eqnarray}
    S_{long}(r)= \frac{r^2}{96 \pi B^5} \,  \exp(-r/B) \, ,
    \label{tail}
\end{eqnarray}

We can determine optimal parameters~$\lambda$ and $B$ by fitting Eqs.~\eqref{eq:long} and ~\eqref{tail} to the source restored from the data, or by fitting the correlation obtained from~\eqref{eq:long} to the measured correlation~\cite{VerdeBuu}. In practice, the results are equivalent in both cases. We obtain the values of $\lambda = 0.38$ and $B=4.0 \, \text{fm}$. The corresponding correlation function and the source function are shown in Fig.~\ref{fig:SF_corr} as blue dashed lines. As seen in panel (a), the experimental data are well reproduced. 

\begin{figure}[htb]
    \centering
    \includegraphics[scale=0.35]{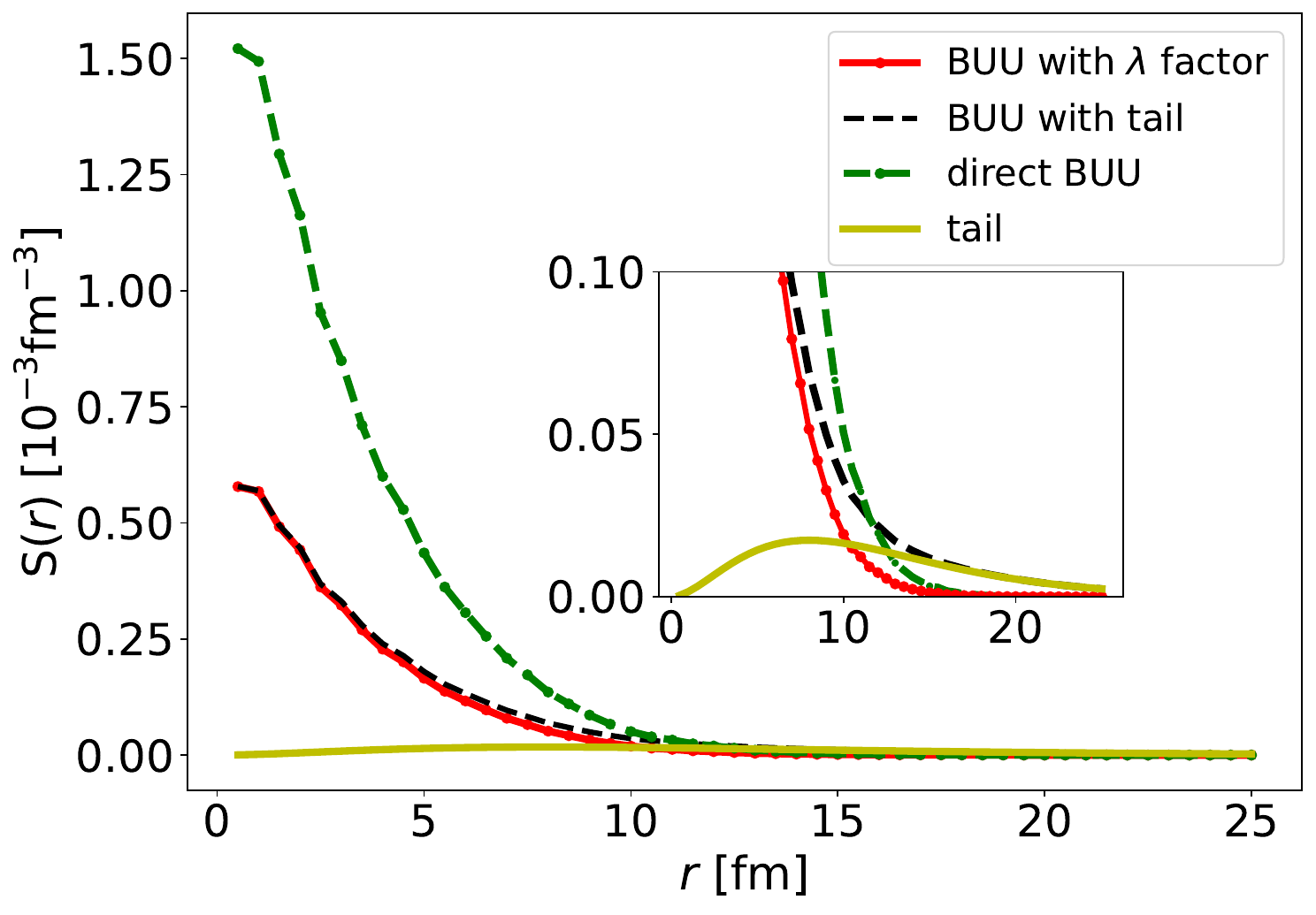}
    \caption{Theoretical $p$--$p$ source function in the $^{36}\mathrm{Ar} + ^{45}\mathrm{Sc}$ reaction at E/A=80 MeV. The green dash-dot line represents the source function obtained directly from pBUU. The red solid line superposed with dots represents the pBUU source multiplied by the $\lambda =0.38$ factor, i.e., the first r.h.s.\ term of Eq.~\eqref{eq:long}. The black dashed line represents the source function obtained using Eq.~\eqref{eq:long}. The yellow solid line shows the separated long-range contribution, i.e., the second term of Eq.~\eqref{eq:long} for $B=4$ fm. The inset shows details in the fall-off of the theoretical sources.}
    \label{fig:SF_tail}
\end{figure}

Although Fig.~\ref{fig:SF_corr} allows for a direct comparison between the measured and predicted correlation functions, it is useful to separate the different source contributions entering Eq.~\eqref{tail} in order to better visualize the role of the additional long-range component. To emphasize the differences between the source contributions, we plot them separately from the deblurred sources in Fig.~\ref{fig:SF_tail}. 
The figure shows the theoretical $pp$ source function for the reaction $^{36}\mathrm{Ar} + ^{45}\mathrm{Sc}$ at a beam energy of 80 MeV/nucleon and an impact parameter of $b=1.9$ fm. The green dashed-dotted line represents the source function obtained directly from pBUU. The solid red line superposed with dots represents the source of pBUU multiplied by the factor $\lambda = 0.38$. The blue dashed line represents the source function of Eq.~\eqref{eq:long}, combining the reduced pBUU source with the additional tail component. The yellow solid line shows the tail function contribution alone, i.e., the second term of Eq.~\eqref{eq:long}, for $B=4$ fm. The inset highlights the fall-off of the theoretical source functions at large relative distances. The contribution of the tail component becomes evident for $r \gtrsim 5$ fm by comparing the blue dashed line (pBUU with tail) with the red solid line superposed with dots (renormalized pBUU source). The difference between these two source functions provides an estimate of the late proton emission contribution to pairs that is not accounted for in the pBUU transport calculation.

Having established the need for a phenomenological long-range source component to explain the correlations in angle-averaged analyses, we next investigate whether such a component may depend on the angular region from which the proton pairs are detected.

\begin{figure*}[htb]
    \centering
    \includegraphics[scale=0.35]{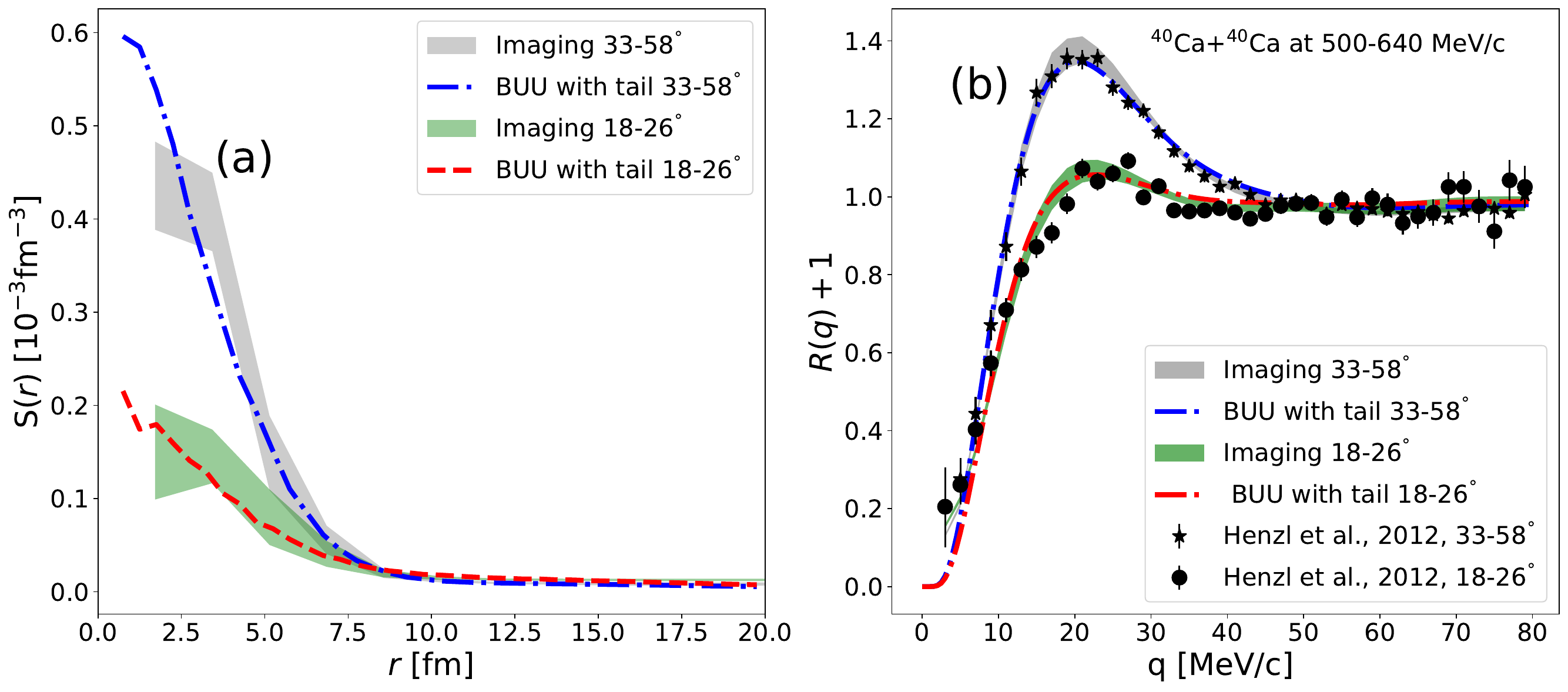}
    \caption{Panel (a) displays the proton–proton source functions for $^{40}\mathrm{Ca} + ^{40}\mathrm{Ca}$ central collisions at $80 \, \text{MeV/nucleon}$, within the total momentum range $500$–$640 \, \text{MeV}/c$ and two angular ranges. The gray band and the blue dash–dotted line represent the source function obtained using the deblurring method and the BUU source supplemented with the long-range supplement, Eq.~\eqref{eq:long}, respectively, for the angular range $33 < \theta \leq 58^{\circ}$. Similarly, the green band and the red dashed line correspond to the deblurring method and the BUU source with the long-range supplement, respectively, for the angular range $18 < \theta \leq 26^{\circ}$. Panel~(b) shows the corresponding correlations. Experimental correlations by Henzl \emph{et al.}~\cite{Henzl} are displayed as stars for $33 < \theta \leq 58^{\circ}$ and as dots for $18 < \theta \leq 26^{\circ}$. Both deblurring (shaded regions) and BUU with a supplement (lines) reproduce the measured correlations well.}
        \label{Ca40CF}
\end{figure*}

\begin{figure*}[htb]
    \centering
    \includegraphics[scale=0.35]{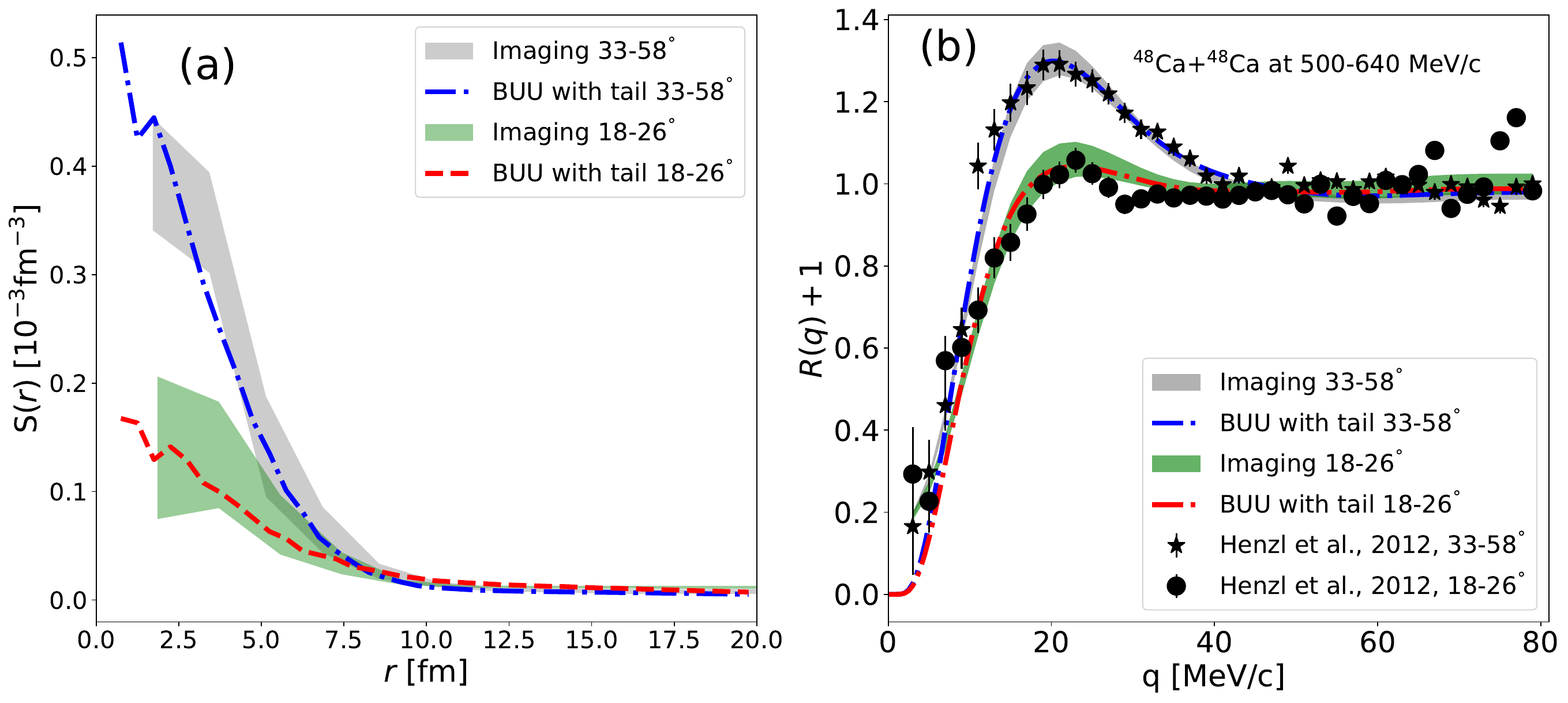}
    \caption{ Similar to Fig.~\ref{Ca40CF}, but now for $^{48}\mathrm{Ca} + ^{48}\mathrm{Ca}$ collisions.} 
        \label{Ca48CF}
\end{figure*}
\section{ Dependence of late proton emission on kinematic region}
\label{sect5} 
In this section, we use measured proton–proton correlation functions to investigate the dependence of late-emission contributions to the relative source on the kinematic region of proton detection. Specifically, we analyze the correlations reported by Henzl et al.~\cite{Henzl} for $^{40}\mathrm{Ca} + ^{40}\mathrm{Ca}$ and $^{48}\mathrm{Ca} + ^{48}\mathrm{Ca}$ collisions at $80~\text{MeV/nucleon}$, with a total momentum cut of 500–640~MeV/$c$. For that momentum cut, the authors reported results from different angular ranges including $18^\circ < \theta \leq 26^\circ$ (forward angles) and $33^\circ < \theta \leq 58^\circ$ (backward angles) for both reactions.  The results from the backward region were already shown in Fig.~\ref{fig:CF_fun}, but not yet analyzed in the context of deblurring and source modification. The region of forward angles in these reactions is dominated by emission from projectile remnants, while backward angles are dominated by participant emission.

\begin{table*}[htb]
   \caption{ The best-fit parameters, $\lambda$ and $B$, from Eqs.~\eqref{eq:long} and \eqref{tail}, are presented here for sources explaining the $pp$ correlations measured in $^{40}\mathrm{Ca} + ^{40}\mathrm{Ca}$ and $^{48}\mathrm{Ca} + ^{48}\mathrm{Ca}$ collisions at $E/A = 80$~MeV for pairs from the total momentum bracket of 500–640~MeV/$c$. We also list the parameters used in the deblurring process, specifically the regularization factor (see \eqref{eq:Lambda}) and the relative momentum resolution in the kernel, $\sigma_q$. }
    \label{tab:my_label}
\centering
\begin{tabular}{|l|c|c|c|c|c|c|r|}
\hline
\hline
\multicolumn{4}{|c|}{Modified source parameters} & \multicolumn{2}{c|}{Deblurring parameters} \\ 
\hline
Reaction& Angular range & $\lambda$ & $B$[fm]  
&Regularization ($\Lambda _0$)& $\sigma _q$ [MeV/c] \\ 
     \hline
    $^{40}\mathrm{Ca} + ^{40}\mathrm{Ca}$& $18 < \theta \leq 26^{\circ}$ &0.15 &4.8 
    &0.0006&2.5\\
    & $33 < \theta \leq 58^{\circ}$ &0.28 &5.2  
    &0.0006&1.5\\
    \hline
    $^{48}\mathrm{Ca} + ^{48}\mathrm{Ca}$&$18 < \theta \leq 26^{\circ}$ & 0.17 & 4.9 
    &0.006&3\\
    & $33 < \theta \leq 58^{\circ}$ & 0.26 &5.5 
    &0.0008&2.5 \\
       \hline
\bottomrule
\end{tabular}
\end{table*}

We have extracted the source functions from the correlations measured within the two angular ranges in the two reactions using the imaging method encapsulated in Eq.~\eqref{eq:RL}. We determined sources in pBUU, two of which are shown in the left panels of Fig.~\ref{fig:CF_fun}.  We then fitted the measured correlations with correlations calculated using the parameterized source, Eqs.~\eqref{eq:long} and \eqref{tail}. The resulting fit parameters are listed in Table~\ref{tab:my_label}, together with the technical parameters in the source imaging. The resolution $\sigma_q$ seems to be better for reaction $^{40}\mathrm{Ca} + ^{40}\mathrm{Ca}$ than $^{48}\mathrm{Ca} + ^{48}\mathrm{Ca}$. The superposition sources, as lines, and the imaged sources, as shaded regions, are shown in the left panels of Figs.~\ref{Ca40CF} and \ref{Ca48CF}, for the collisions $^{40}\mathrm{Ca} + ^{40}\mathrm{Ca}$ and $^{48}\mathrm{Ca} + ^{48}\mathrm{Ca}$, respectively.  In the right panels, the corresponding correlation functions for the two reactions, related to the sources via the KP equation \eqref{KPeqn_iso}, are shown, together with the data~\cite{Henzl}.  The data description is adequate in each case. 

It is important to note that for the results obtained through imaging here, we employ a regularization factor in the RL algorithm depending on the relative distance: 
\begin{equation}
    \Lambda = \Lambda_0\left(\frac{r}{R_0}\right)^2,
    \label{eq:Lambda}
\end{equation}
where the values of the constant $\Lambda_0$ are shown in Table~\ref{tab:my_label}, and we use $R_0 = 10$~fm.  With the form \eqref{eq:Lambda}, we prevent short-wavelength oscillations in the source at far distances, on which the measured correlations provide very limited information.

\begin{figure}[htb]
    \centering
    \includegraphics[scale=0.35]{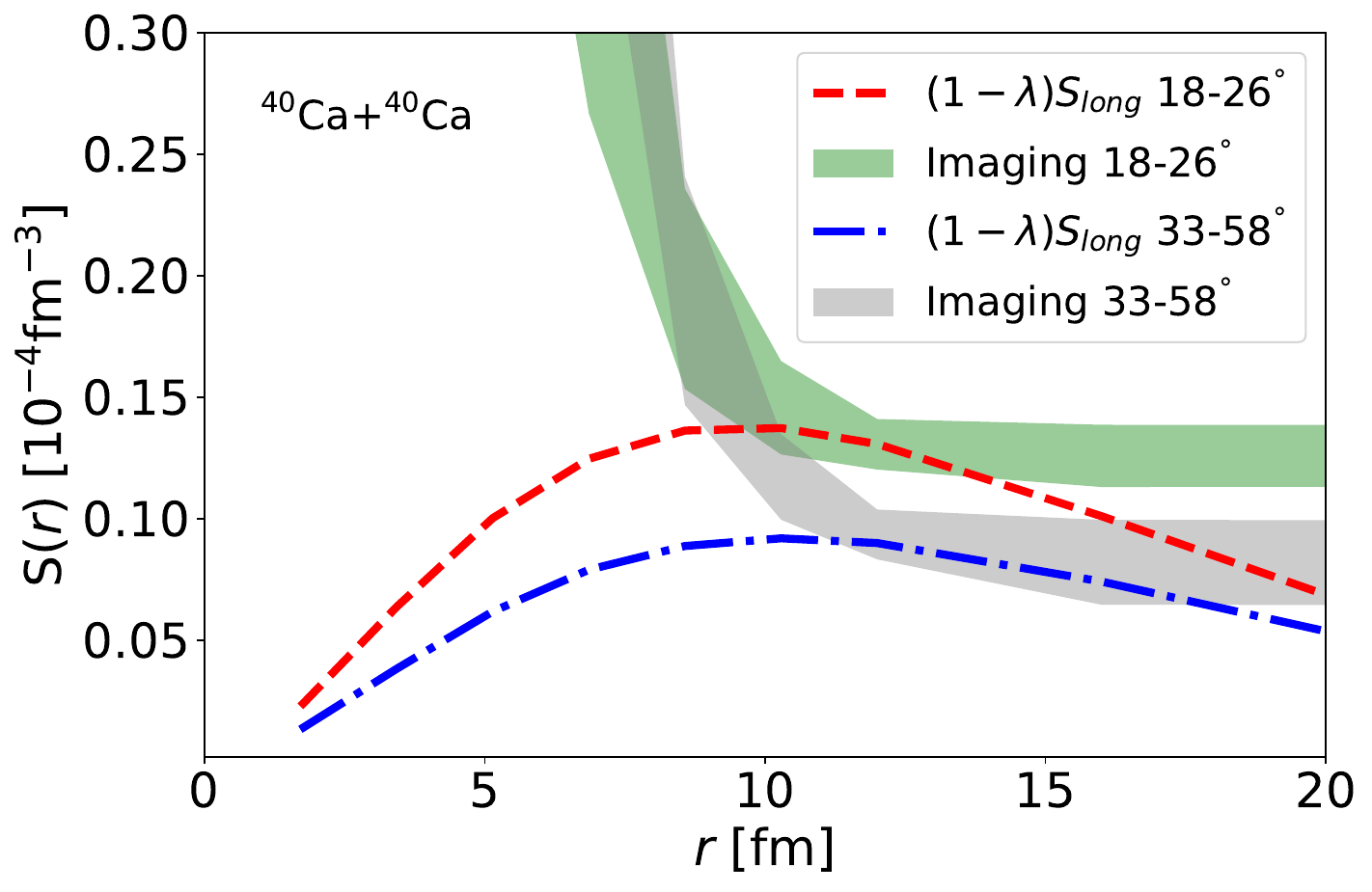}
    \caption{Parametrized long-range source contributions, $(1-\lambda)\, S_{long}(r)$, for forward ($18 < \theta \leq 26^{\circ}$) and backward ($33 < \theta \leq 58^{\circ}$) angles, represented with dashed (red) and dot-dashed (blue) lines, respectively, in the central $^{40}\mathrm{Ca}+^{40}\mathrm{Ca}$ collisions at $80 \, \text{MeV/nucleon}$ and total momentum region of $500$–$640 \, \text{MeV}/c$~\cite{Henzl}.  The parameter values for the long-range sources can be found in Table \eqref{tab:my_label}.  Alongside the long-range source contributions, the sources imaged directly from the correlation functions are shown for the forward and backward angles as green- and gray-shaded regions, respectively, with the width of each region at any fixed distance representing the uncertainty.}
        \label{C40tal}
\end{figure}

\begin{figure}[htb]
    \centering
    \includegraphics[scale=0.35]{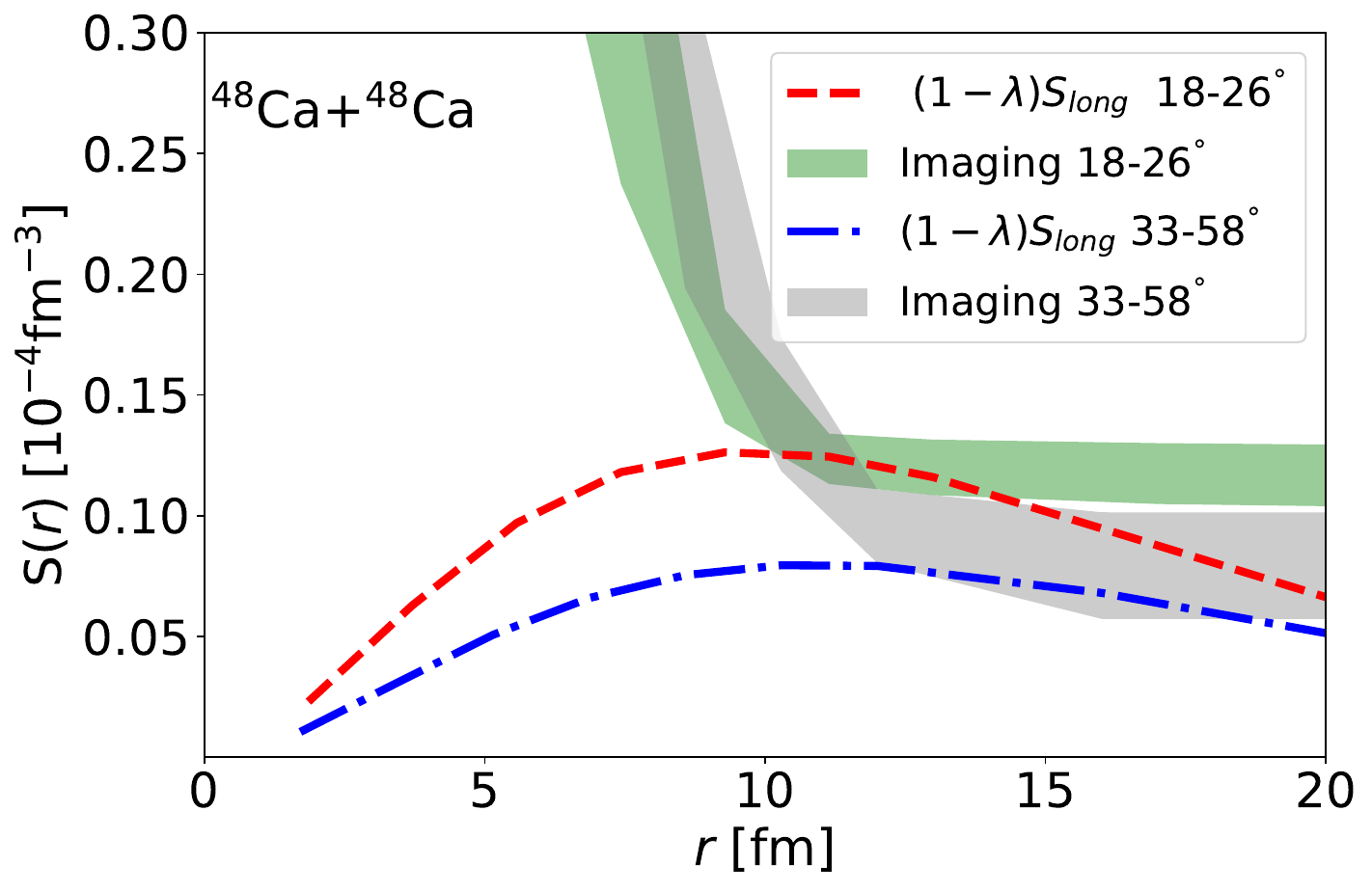}
    \caption{Analogous content to that in Fig.~\ref{C40tal}, but now for $^{48}\mathrm{Ca} + ^{48}\mathrm{Ca}$ collisions.}
    
    \label{C48tail}
\end{figure}


One remarkable feature of the results in Table~\ref{tab:my_label} and earlier is the low values of $\lambda$, no matter which kinematic region is chosen at the given beam energy of 80 MeV/nucleon.  This indicates that late emissions make a strong contribution to the measured pairs.  It should be noted that a single late-emitted proton in the pair relegates the pair to late emission. I.e., $\lambda = 0.25$ should be understood as about 50\% of protons in a given kinematic region that stem from fast emission and the other 50\% from slow emission.  The higher values of the source at short distances in Figs.~\ref{Ca40CF} and \ref{Ca48CF} for backward angles than for forward angles, tied to higher values of $\lambda$, indicate a stronger contribution of direct emission relative to secondary emission to backward angles, as physically expected. Already, the more pronounced peak in the correlation function at $q \approx 20 \, \text{MeV}/c$ for backward angles than forward angles indicates that.  The peak is related to the resonance enhancement of the probability density at short distances in the kernel $K$ in the KP equation~\cite{nzabahimana2023source}.  The backward-angle source is more intense at short distances in the collision $^{40}\mathrm{Ca} + ^{40}\mathrm{Ca}$ than in the collision $^{48}\mathrm{Ca} + ^{48}\mathrm{Ca}$, consistent with the smaller system.  Notable is that the slope parameter is about the same, $B \approx 5 \, \text{fm}$, for the fitted long-range part source that complements the transport sources in Table~\ref{tab:my_label}, regardless of which system and angular range is considered.  More on the supplementing sources below.

In Figs.~\ref{Ca40CF} and \ref{Ca48CF}, we observe a reasonable agreement between the imaged and transport sources supplemented with the tails.  Notably, imaging cannot provide good resolution for source values at short distances, as the contribution of the source to the correlation function is weighted by an $ r^2$ factor.  The parametrized tail additions appear to also bring about agreement with the imaged sources at larger distances.  However, we provide more insight into the sources at larger distances in Figs.~\ref{C40tal} and~\ref{C48tail}.  Specifically, we show there the tail additions to the transport sources for different angles together with the imaged sources.  The stronger intensity of the additions for forward than backward angles is apparent.  In terms of range, these additions dominate the combined sources past $r \approx 10 \, \text{fm}$, as can be inferred in Fig.~\ref{fig:SF_tail} by comparing the BUU contribution scaled down with the factor $\lambda$ and the tail addition.  In Figs.~\ref{Ca40CF} and~\ref{Ca48CF}, it is found that the tail function improves the agreement with the imaged sources at least up to a separation of $\approx 15 \, \text{fm}$, but then the parametrized sources fall below the imaged ones. The discrepancy at the still larger distances may be tied to the discrepancy between the correlation functions from the combined sources and the data at the lowest $q$ values in the measurements shown in Figs.~\ref{Ca40CF} and~\ref{Ca48CF}.  With more details available for low correlations $q$, additional parameters in the long-range source contribution~\eqref{eq:long} may be called for.

\section{Summary and outlook \label{sum}}

In this work, we have confronted the proton-proton correlation functions measured in the reactions $^{36}$Ar $+^{45}$Sc, for $^{40}\mathrm{Ca} + ^{40}\mathrm{Ca}$ and $^{48}\mathrm{Ca} + ^{48}\mathrm{Ca}$ at E/A = 80~MeV with the predictions of the pBUU transport model. In our calculations, the shape of the pBUU correlation functions for this system shows very weak sensitivity to the stiffness of the EoS and to momentum-dependent interactions. As previously observed in the literature, the absence of secondary decay modeling in transport models makes it difficult to reproduce the experimental correlation function directly. We used a deblurring imaging technique to invert the KP equation and determine the source, thereby better revealing the effects of long-lived emissions. Then, this source shape was compared with pBUU-calculated sources and used to assess the secondary-decay contributions to the source more quantitatively than in the past. 
A renormalization of a transport-model source function, employed previously in the literature, while forcing it to fit the height of the peak at 20 MeV/c in the correlation function, cannot address the correlation function shape at small relative momenta.  Instead, adding an analytical tail function to the pBUU emitting source and fitting its two free parameters to the deblurred source, leads to a two-proton source function that adequately describes the experimental correlation function at small $q$-values and the peak height. Given that only a few data points may be obtained in measurements at low relative $q$, the falloff scale for the source tail at intermediate distances may be all that can be firmly established.  However, as measurement resolution improves, more may be learned from imaging and/or by expanding source parameterization.  From the perspective of transport models, the simple parametrization introduced for the source function allows quantification of shortcomings of a transport model, such as pBUU, that specific or other models may eventually need to address.

Finally, we studied the angular dependence on the slow proton emission for $^{40}\mathrm{Ca} + ^{40}\mathrm{Ca}$ and $^{48}\mathrm{Ca} + ^{48}\mathrm{Ca}$ collisions at $80~\text{MeV/nucleon}$, with a total momentum cut of 500–640~MeV/$c$. We found a greater contribution of slow proton emission to forward than backward angles in this experiment.

Looking forward, it would be important to study directional correlation functions by expanding the measured functions and pBUU source functions into angular components~\cite{tesseral}. Especially, using the quadrupole components of the source function may help us gaining more quantitative information about the distribution of emissions from second decays. In addition, a systematic Femtoscopy analysis, extended to various heavy-ion collision systems across ranges of mass and beam energy, may help improve transport models and provide useful probes of the nuclear equation of state.
\section*{Acknowledgments}
This work was supported by the U.S. Department of Energy through Los Alamos National Laboratory. Triad National Security, LLC operates Los Alamos National Laboratory for the National Nuclear Security Administration of the U.S.\ Department of Energy~(Contract No.\ 89233218CNA000001). The authors acknowledge the support from the US Department of Energy Office of Science under Grant No.\ DE-SC0019209.
\nocite{}
\bibliographystyle{apsrev}
\bibliography{reference.bib}
\end{document}

%% file: abstract.tex
Two-particle correlations play a pivotal role in understanding the space-time characteristics of particle emission in Heavy-Ion Collisions (HIC). A relative emission source typically represents these characteristics and can be obtained from transport model simulations such as the Boltzmann-Uehling-Uhlenbeck (BUU) transport model.  In this paper, we use the BUU transport model to simulate p-p sources while simultaneously using HIC correlation data to infer them for the same collisions, combining the Koonin-Pratt equation with Richardson-Lucy deblurring. In comparing the sources, we can isolate and provide a simple parametrization for source tails missing from BUU sources.  These tails may be attributed to secondary decays that contribute at least one member of the pair in the source.  With a tail added, the BUU can reproduce 80 MeV/u Ar + Sc and Ca + Ca measurements.  We investigate the angular dependence of Ca + Ca correlations to infer the angle dependence of slow emissions.  Our parametrization for the tail may be employed when no deblurring for source inference is pursued.